\documentclass[aps,prl,twocolumn,showpacs,floatfix,nofootinbib]{revtex4}
 \usepackage{float,ulem}
 \usepackage{graphicx,epsfig,epstopdf,amsmath}

\newcommand{\be}{\begin{equation}}
\newcommand{\ee}{\end{equation}}
\newcommand{\nn}{\nonumber\\}

\begin{document}

%\begin{frontmatter}

\title{Thermal and viscous dissipation in relativistic heavy ion collisions}
\author{Sukanya Mitra}
\email{sukanya.mitra@tifr.res.in}
\author{Subrata Pal}
\address{Department of Nuclear and Atomic Physics, Tata Institute of Fundamental Research, 
Homi Bhabha Road, Mumbai 400005, India}

\begin{abstract}

We investigate the effects of finite baryon density and temperature on the bulk properties of matter 
formed in relativistic heavy ion collisions within second-order dissipative hydrodynamics.
The relativistic fluid evolution equations for heat flow and shear stress tensor are derived from
kinetic theory by using Grad's 14-moment approximation for the single-particle phase-space distribution function.
The new equations provide a number of additional terms associated with heat-shear couplings as compared to 
the existing derivations based on entropy principle. The dissipative equations are encoded in 
non-boost-invariant hydrodynamic model simulation and studied for the evolution of high baryon density
matter encountered at the beam energy scan program at RHIC. We find that thermal dissipation dominates 
shear pressure in defining the bulk observables at the low energy but its effect diminishes
at ultra-relativistic energies.

\end{abstract}

%\pacs{25.75.Ld, 24.10.Nz, 47.75+f}
% 25.75.Ld Collective flow
% 24.10.Nz Hydrodynamic models
% 47.75.+f Relativistic fluid dynamics
% 47.10.ad Navier-Stokes equations

%\end{frontmatter}
\maketitle

\section{Introduction}

Heavy-ion collision experiments at the Relativistic Heavy Ion Collider (RHIC) energy of $\sqrt{s_{NN}} = 200$ GeV 
\cite{Adams:2005dq,Adcox:2004mh} and at the Large Hadron Collider (LHC) energy $\sqrt{s_{NN}} = 2.76$ TeV 
\cite{ALICE:2011ab,ATLAS:2012at,Chatrchyan:2013kba} have already provided conclusive evidence of 
the formation of a strongly interacting QCD matter at vanishing net-baryon density. Such a conclusion is based on the 
relativistic viscous hydrodynamic analysis of the observed strong collective flow that require a small shear viscosity to entropy 
density ratio as well as a lattice QCD equation of state at zero chemical potential that predicts a smooth 
crossover from hadron to quark phase at a temperature of $T_c \approx 154$ MeV \cite{Heinz:2013th}.

While finite chemical potential lattice QCD calculation is notoriously difficult due to the sign problem,
a detailed study of the transport properties of matter at finite baryon density and temperature
could provide valuable information on nuclear phase diagram which is expected to be first order.
The ongoing beam energy scan (BES) program at RHIC, the NA61/SHINE experiment at SPS and the 
future FAIR facility at GSI are all dedicated to explore the hot and dense matter formed
at midrapidity.

Second-order relativistic hydrodynamic theories 
have been quite successful in describing the final-state observables
at RHIC and LHC where the system is expected to thermalize \cite{DNMR,Gale:2013da,Bhalerao:2015iya}.
Since hydrodynamics is an effective macroscopic theory based on gradient expansion of thermodynamic 
state variables up to certain order \cite{Degroot}, it is expected to break down in systems with very 
large spatial and/or temporal gradients. However, the effectiveness in explaining collisions involving small systems 
such as protons and light nuclei, where the medium is not expected to thermalize, have extended the applicability of 
hydrodynamics to the far-from-equilibrium domain as can be envisaged at the lower collision energies 
\cite{Heller:2015dha,Romatschke:2017vte}.

Most of the hydrodynamic analysis of heavy/light-ion data so far has been confined at ultra-relativistic energies
in the boost invariant central rapidity region where the heat transport is ignored as compared to viscous dissipation. 
A sizable thermal dissipation may arise from large spatial gradients in the chemical potential and temperature
at the lower collision energies due to baryon stopping (at midrapidity) as well as at the higher rapidities
due to the presence of nuclear spectators. Only a few hydrodynamic calculations with finite baryon density
exist that are either derived from entropy maximization and lacks the microscopic dynamics
found in kinetic theory \cite{MurongaRischke,Betz,Huovinen,Bouras}, or ignores the coupling between thermal and 
viscous evolution \cite{Li-Shen}.

In this article we investigate the impact of both heat and viscous transport on bulk dynamics 
in heavy-ion collision over an energy range $\sqrt{s_{NN}} = 4.8 - 200$ GeV.
We present a new derivation of viscous and heat dissipation equations in second-order hydrodynamics 
using Grad's 14-moment \cite{Grad} method.  We find a number of additional terms compared to the traditional 
Muller-Israel-Stewart formalism \cite{MIS} due to the heat-viscous coupling.
The implications of these dissipative equations have been demonstrated within a non-boost-invariant longitudinal
expansion of matter \cite{Chattopadhyay:2018dth}. Noteworthy of which is that at lower collision energies, heat 
dissipation enhances and becomes comparable to viscous effects.

\section{Relativistic hydrodynamics}

The conserved particle four-current and the energy-momentum tensor can be expressed \cite{Degroot}
in terms of single particle phase space distribution $f_p \equiv f(x,p)$ as 
\begin{align}
N^\mu &  = g\int dp \ p^\mu (f_p - \bar f_p) =  n u^\mu + n^\mu , \label{field1}\\
T^{\mu\nu} & = g\int dp \ p^\mu p^\nu (f_p  + \bar f_p) = \epsilon u^\mu u^\nu - (P+\Pi)\Delta^{\mu\nu} \nonumber \\
& ~~~~~ + \pi^{\mu\nu} + \left( q^\mu + W^\mu \right) u^\nu  + \left( q^\nu + W^\nu \right) u^\mu , \label{field2}
%% [ q^\nu + (\epsilon+P)\Delta^{\nu\sigma} N_\sigma ] u^\mu , \label{field2}
\end{align}
where the phase-space factor is
$dp = d{\bf p}/[(2\pi)^3\sqrt{{\bf p}^2+m^2}]$ for a particle of rest mass $m$, degeneracy $g$ with
four-momentum $p^\mu$ in a system composed of particles of single species. In the above tensor decomposition,
$(n,\epsilon,P)$ are the net-particle number density, energy density and pressure density. 
$\Delta^{\mu\nu} = g^{\mu\nu}-u^\mu u^\nu$ is projection operator orthogonal to the hydrodynamic four-velocity
$u^\mu$. The dissipative quantities are the charge diffusion current $n^\mu \equiv nW^\mu/(\epsilon+P) = W^\mu/h$, 
the bulk viscous pressure
$\Pi$, the shear stress tensor $\pi^{\mu\nu}$, and the heat flow $q^\mu$ defined as 
$q^\mu = u_\nu T^{\nu\sigma}\Delta_{\sigma}^{\mu}-hN^{\sigma}\Delta_{\sigma}^{\mu}$.
We have used Eckart's choice of velocity frame where 
$n^\mu=\Delta^{\mu\nu}N_{\nu}=0$ and the heat flow 
$q^\mu= u_\nu T^{\mu\sigma}\Delta^{\mu}_{\sigma}$. 

The fundamental conservation equations of particle current, $\partial_{\mu}N^{\mu}=0$, and
energy-momentum tensor $\partial_{\mu}T^{\mu\nu}=0$ give the evolution equations for $n$, $\epsilon$ and $u^\mu$
\begin{align}
& \dot{n} + n\theta = 0~, \label{thid1}\\
& \dot{\epsilon} + (\epsilon+P+\Pi)\theta - \pi^{\mu\nu}\sigma_{\mu\nu} - 2q^{\mu}\dot{u}_{\mu} + \nabla_{\mu}q^{\mu} = 0 ,\label{thid3}\\
& (\epsilon+ P +\Pi)\dot{u}^\alpha - \nabla^{\alpha}(P+\Pi) + \Delta^{\alpha}_{\nu}\partial_{\mu}\pi^{\mu\nu} 
 - \pi^{\alpha\beta} \dot{u}_\beta   \nonumber  \\
& + \Delta^{\alpha}_{\nu}\dot{q}^{\nu} + q^{\alpha}\nabla_{\nu}u^{\nu}  
  + q^{\nu}\nabla_{\nu}u^{\alpha}  = 0 . \label{thid2} 
 \end{align}
We have used the standard notation 
$\dot{A} \equiv u^\mu\partial_\mu A$ for comoving derivatives, 
$\nabla^\mu \equiv \Delta^{\mu\nu}\partial_\nu$ for space-like derivatives, 
$\theta \equiv \partial_\mu u^\mu$ for the expansion scalar,
and $\sigma^{\mu\nu} = \frac{1}{2}(\nabla^\mu u^\nu + \nabla^\nu u^\mu)- \frac{1}{3}\theta \Delta^{\mu\nu}$ for the
velocity stress tensor. In the present calculation we have ignored the effects of bulk viscosity by setting $\Pi=0$.

For a system close to local thermodynamic equilibrium, $f_p$ can be written as 
$f_p = f^0_p + \delta f_p$.
The equilibrium distribution function is defined as $f^0_p = [\exp(\beta u \cdot p - \alpha) + r]^{-1}$ 
(with $r =1,-1,0$ for Fermi, Bose, Boltzmann gas) where
$\beta=1/T$ is the inverse temperature, $\mu$ the chemical potential with $\alpha = \beta\mu$, and
the scalar product $u \cdot p \equiv u_\mu p^\mu$. From Eqs. (\ref{field1}) and (\ref{field2}), the
dissipative quantities can then be expressed in terms of $\delta f_p$ as 
\begin{align}
& n^\mu  =  \Delta_\alpha^\mu \: g \int dp \  p^\alpha(\delta f_p - \delta\bar f_p) , \label{curnt} \\
& \pi^{\mu\nu}  =  \Delta_{\alpha\beta}^{\mu\nu}  \: g \int dp \ p^\alpha p^\beta (\delta f_p + \delta\bar f_p) , \label{shear} \\
& q^\mu  =  \Delta^{\mu\nu} \: g \int  dp \  p_\nu \Big[ p^\alpha u_\alpha (\delta f_p + \delta\bar f_p) 
 - \frac{\epsilon + P}{n}(\delta f_p - \delta\bar f_p) \Big], \label{heat} 
\end{align}
where $\Delta^{\mu\nu}_{\alpha\beta} \equiv 
\frac{1}{2}(\Delta^\mu_\alpha \Delta^\nu_\beta + \Delta^\mu_\beta \Delta^\nu_\alpha)
- \frac{1}{3} \Delta^{\mu\nu} \Delta_{\alpha\beta}$ is traceless projection operator orthogonal to 
$u_\mu$ and $\Delta_{\mu\nu}$.

\section{Dissipative evolution equations}

To derive the dissipative evolution equations we require the out-of-equilibrium distribution function
$\delta f_p$. This can be obtained by using the relativistic Boltzmann transport equation 
$p^\mu \partial_\mu f(x,p)= C[f]$, and recasting $\delta f_p$ 
as $\delta f_p=f^0_p(1\pm f^0_p)\phi_p$ with $\phi_p$ being the deviation
function. The linearized Boltzmann equation can be then written as \cite{Degroot}
\be
\Pi^{\mu}\partial_{\mu}f^0_p + f_p^0(1\pm f_p^0)\Pi^{\mu}\partial_{\mu}\phi_p
+ \phi_p \Pi^{\mu}\partial_{\mu}f_p^0 =-\beta{\cal L}_p[\phi_p] ,
\label{secondhydro0}
\ee
with $\Pi^{\mu}=\beta p^\mu$ is the scaled particle four-momenta and 
$\tau_p=\beta p^{\mu}u_{\mu}$ is the scaled energy (used below) in the local rest frame. 
The linearized collision operator is then given by
\begin{align}
{\cal L}_{p}[\phi_p]= g & \int dp' dk \: dk' \ f_p^0 (1 \pm f_k^{0}) f_{p'}^0(1\pm f_{k'}^{0})\nonumber\\
& \times [\phi_{p}+\phi_{p'}-\phi_{k}-\phi_{k'}] W_{pp' \to kk'} ,
\label{coll2} 
\end{align}
where $W_{pp' \to kk'}$ is the transition rate.
To obtain $\phi_p$ using the Eq. (\ref{secondhydro0}), we take recourse to Grad's 14-moment
method \cite{Grad} for $f_p$ in orthogonal basis.
In this approach, the scalar $\phi_p$ is expanded in the particle-momentum space, 
and expressing it in terms of scalar products 
of tensors formed from $p^{\mu}$ and tensor functions of $x_{\mu}$ as \cite{Mitra}
\be
\phi_p=B_p^{\mu}(x,\tau_p)\langle\Pi_{\mu}\rangle - C_p^{\mu\nu}(x,\tau_p)\langle\Pi_{\mu}\Pi_{\nu}\rangle ,
\label{Gradphi}
\ee
where the irreducible tensors are  $\langle\Pi_{\mu}\rangle = \Delta_{\mu\nu} \Pi^\nu$
and $\langle\Pi_{\mu}\Pi_{\nu}\rangle = \Delta^{\alpha\beta}_{\mu\nu} \Pi_\alpha \Pi_\beta$.
The coefficients $B_p^{\mu}$ and $C_p^{\mu\nu}$ are
further expanded in a power series of $\tau_p$ as:
\be
B_p^{\mu}=\sum_{s=0}^{1} [B_s(x)]^{\mu} \: \tau_p^s ,~~~~
C_p^{\mu\nu}=\sum_{s=0}^{0}[C_s(x)]^{\mu\nu} \: \tau_p^s.
\label{Gradcoeff}
\ee
Here the polynomials, up to first non-vanishing contribution to irreversible flows, have been retained. 
The unknown coefficients $B_s$ and $C_s$ can be obtained in form of the 
dissipative fluxes by putting Eq. (\ref{Gradphi}) in Eqs. (\ref{curnt})-(\ref{heat}).
By using the the moment integrals,
\be
F_p^{\nu_1\cdots\nu_n}=\int dp \ f_p^0(1\pm f_p^0)p^{\nu_1\cdots\nu_n}=\sum_{l=0}^{[n/2]}a_{nl}(\Delta u)_{nl},
\label{momin}
\ee
\begin{align}
{\rm with}~~ a_{nl} & = (-1)^l\frac{^{n}C_{2l}}{2l+1}\int dp \mid\vec{p}\mid^{2l}(p^0)^{n-2l}f_p^0(1\pm f_p^0),\nonumber\\
(\Delta u_{nl}) & =\frac{1}{n!}\sum_{\rm perm}\Delta^{\nu_1\nu_2}\cdots\Delta^{\nu_{2l-1}\nu_{2l}}u^{\nu_{2l+1}}\cdots u^{\nu_n},
\end{align}
we obtain the coefficients
\begin{align}
&C_0^{\mu\nu}  = \kappa_0 \pi^{\mu\nu},  ~~~\kappa_0 = -[T^2 c_0]^{-1}, \label{coeff1} \\
&B_1^\mu  = \beta_1 q^\mu,  ~~~~~~\beta_1 = [T^2(b_2 - \hat{h} b_1)]^{-1},  \label{coeff2}
\end{align}
where $\hat{h} = w/nT$ with $w=\epsilon+P$ being the enthalpy of the system.
Further, using Eckart's definition of velocity, the coefficients $B_0^\mu$ and $B_1^\mu$ can be related as 
\be
B_0^{\mu}=-B_1^{\mu} b_1/b_0 = \beta_0 q^{\mu}, ~~ \beta_0 = -\beta_1 b_1/b_0.
\label{coeff3}
\ee
The quantities $b_s$ and $c_s$ are actually the moment integrals defined as 
\begin{align}
\Delta^{\mu\nu}b_s & = g\int dp \ f_k^0(1\pm f_p^0)\langle\Pi^{\mu}\rangle\langle\Pi^{\nu}\rangle \tau_p^s ,\\
\Delta^{\alpha\beta\mu\nu}c_s & = g\int dp \ f_p^0(1\pm f_p^0) 
\langle\Pi^{\alpha}\Pi^{\beta}\rangle\langle\Pi^{\mu}\Pi^{\nu}\rangle \tau_p^s .
\end{align}
Equations (\ref{coeff1}), (\ref{coeff2}), (\ref{coeff3}), give the three unknown coefficients necessary 
to specify the out-of-equilibrium distribution from (\ref{Gradphi}).
With the help of moment integrals (\ref{momin}), the deviation function becomes
\be
\phi_p=\frac{q^\mu \langle\Pi_{\mu}\rangle}{w} \frac{n(p^\nu u_\nu - 4T)}{w - 5nT}
+\frac{\pi^{\mu\nu} \langle\Pi_{\mu}\Pi_{\nu}\rangle}{2w} .
\label{Gradphi1}
\ee
Using Eqs. (\ref{secondhydro0}) and (\ref{Gradphi1}) along with the moment integrals, we finally
obtain the evolution equation for the shear stress tensor and heat flow 
\begin{align}
\Delta^{\alpha\beta}_{\mu\nu} \dot{\pi}^{\mu\nu} & = \frac{2\eta_v}{\tau_\pi} \sigma^{\alpha\beta}  
 - \frac{\pi^{\alpha\beta}}{\tau_\pi} \nn
& + r_{\theta} \pi^{\alpha\beta} \partial\cdot u + r_{\omega} \pi_{\rho}^{\langle\alpha}\omega^{\beta\rangle\rho}
+ r_{\sigma} \pi_{\rho}^{\langle\alpha} \sigma^{\beta\rangle\rho} \nn
& + r_a q^{\langle\alpha} \dot{u}^{\beta\rangle} + r_I\nabla^{\langle\alpha}q^{\beta\rangle}
+r_T q^{\langle\alpha}\frac{\nabla^{\beta\rangle}T}{T} , \label{shearhydro0} \\
\Delta^{\alpha}_{\mu} \dot{q}^{\mu} & = \frac{\lambda T}{\tau_q} 
\left(\frac{\nabla^{\alpha}T}{T} - \dot{u}^\alpha \right) - \frac{q^{\alpha}}{\tau_q} \nn
& + l_{\sigma}q_{\mu}\sigma^{\alpha\mu} + l_{\omega}q_{\mu}\omega^{\alpha\mu} 
+ l_{\theta}q^{\alpha}\partial\cdot u  \nn
& + l_a \pi^{\alpha\mu} \dot{u}_{\mu} + l_T \pi^{\alpha\mu} \frac{\nabla_{\mu}T}{T} 
+ l_\pi  \Delta^{\alpha\mu}\nabla^{\nu} \pi_{\mu\nu}.
       \label{heathydro0}
\end{align}
Here $\eta_v$ is the shear viscosity and $\lambda$ is the thermal conductivity of the system.
The chemical potential gradient is converted to temperature and pressure gradients 
by applying the Gibbs' Duhem relation 
$\partial^{\mu}(\mu/T) = (nT)^{-1}\partial^{\mu}P - \hat{h}^{-1}(\partial^{\mu}T)/T$.
The relaxation times for shear pressure $\tau_\pi = 6\eta_v/w$ and 
heat flow $\tau_q = 5\lambda T(\hat{h} - 3)/[w(\hat{h}-5)^2]$, as well as all the 
second-order transport coefficients are explicitly determined in terms of the hydrodynamic variables.
We note that the shear-heat coupling terms obtained in kinetic theory, 
$q^{\langle\alpha}\frac{\nabla^{\beta\rangle}T}{T}$, $~q_{\mu}\sigma^{\alpha\mu}$
and $~\pi^{\alpha\mu} \frac{\nabla_{\mu}T}{T}$ are new compared to previous studies of
thermal diffusion \cite{Huovinen,Bouras} by entropy maximization.
Moreover, the coupled dissipative equations found here are distinct to the decoupled
shear and charge current equations obtained in Chapman-Enskog like iterative approach 
of the Boltzmann equation in the relaxation time approximation \cite{Jaiswal:2015mxa}.
Equations (\ref{shearhydro0}) and (\ref{heathydro0}) along with their set of coefficients 
constitute one of the main results in the present study.

The entropy four-current of the single-component system, 
$S^{\mu}_r=-g\int dp \: p^{\mu} [f_p\ln f_p + r(1 - r f_p) \ln (1 - r f_p)]$, 
can be decomposed into
two parts:  $S^{\mu}=su^{\mu}+\Phi^{\mu}$ with $s=u^{\mu}S_{\mu}$ as the entropy density,
and $\Phi^{\mu}=\Delta^{\mu\nu}S_{\nu}$ as the entropy flux.
Substituting the out-of-equilibrium distribution function from (\ref{Gradphi1}) into $S^{\mu}$ we get,
\begin{align}
s & = s_{eq} + \frac{3q^\mu q_\mu}{T(\epsilon +P)} - \frac{3\pi^{\mu\nu} \pi_{\mu\nu}}{2T(\epsilon +P)},
\label{ent4}\\
\Phi^\mu & =  \frac{q^\mu}{T} +\frac{2\pi^{\mu\nu} q_\nu}{T(\epsilon +P)} ,
\label{ent5}
\end{align}
with the equilibrium entropy density $s_{\rm eq}= (\epsilon + P - \mu n)/T$. 
It is important to note that within second-order dissipative hydrodynamics, 
the entropy flux vanishes in absence of heat flow (see Ref. \cite{Chattopadhyay:2014lya})
and originates entirely due to thermal conduction.

\section{Non-boost-invariant dissipative hydrodynamics}

To demonstrate the numerical significance of viscous and thermal dissipation equations obtained here, 
we consider a non-boost-invariant longitudinal expansion of matter \cite{Chattopadhyay:2018dth}
at finite baryon density.
In terms of the Milne coordinates ($\tau,x,y,\eta$), where $\tau=\sqrt{t^2-z^2}$ and space-time rapidity 
$\eta=\tanh^{-1}(z/t)$, the hydrodynamic four-velocity, $u^{\mu}=\gamma(1,0,0,v_{\eta})$, 
includes a longitudinal component $v_\eta$. 

The components of $\pi^{\mu\nu}$ can be obtained from azimuthal symmetry, orthogonality to $u^\mu$ 
and tracelessness. This leads to only one independent component which we take as $\pi^{\eta\eta}$. 
Similarly, for the heat flow, using the orthogonality condition $u_{\mu}q^{\mu}=0$, the nonvanishing  
components are $q^{\eta}=-\gamma q/\tau$ and $q^{\tau}=-\gamma\tau v_{\eta} q$.
The evolution equations for the (scaled) independent components of shear stress, 
$\pi \equiv -\tau^2\pi^{\eta\eta}/\gamma^2$, and heat flow $q \equiv - \tau q^{\eta}/\gamma$ 
in (1+1)D then become
\begin{align}
D\pi & = \frac{4}{3}\frac{\eta_v}{\tau_\pi}\theta - \frac{\pi}{\tau_\pi} - 2\pi\theta  \nn
& - \frac{2}{3} \tau v_{\eta} \left( r_T q \frac{\nabla^I T}{T}  
+ r_\alpha q \frac{\nabla^I P}{w} + r_I \nabla^{I}q \right) ,
\label{evo1}\\
Dq & = \frac{\lambda T}{\tau_\lambda} \tau v_{\eta} \left( \frac{\nabla^I T}{T}
- \frac{\nabla^I P}{w} \right) - \frac{q}{\tau_\lambda} + \frac{2}{3}(l_\sigma - 2) q\theta \nn
& + \tau v_{\eta} \left( l_T \pi\frac{\nabla^I T}{T} + l_\sigma \pi\frac{\nabla^I P}{w}
+  l_\pi \nabla^I\pi \right).
\label{evo2}
\end{align}
Here $D=\gamma(\partial_\tau + v_\eta \partial_\eta)$ is the time derivative in the local fluid rest frame,
$\theta = \partial \cdot u = \tau^{-1}\partial_\tau (\tau\gamma) + \partial_\eta (\gamma v_\eta)$ is the
local expansion rate and  $\nabla^I =\gamma(\partial_\tau + \tau^{-2}v_{\eta}^{-1}\partial_\eta)$.

Taking the two independent components of $T^{\mu\nu}$ as $T^{\tau\tau}$ and $T^{\tau\eta}$, the
evolution equations (\ref{thid3}) and {\ref{thid2}) for energy density and velocity reduce
in the (1+1)D flow to
\begin{align}
&\partial_\tau \tilde{T}^{\tau\tau} + \partial_\eta (\tilde{v}_\eta \tilde{T}^{\tau\tau})
= -\gamma^2\left(\epsilon + P_L\right) + \epsilon + 2\gamma^2\tau v_{\eta} q , \label{evo3}\\
& \partial_\tau \tilde{T}^{\tau\eta} + \partial_\eta ( v_\eta \tilde{T}^{\tau\eta}
+ \tilde{P}_L/\tau^2 - v_{\eta}q ) \nn
& = - 2\gamma^2 v_{\eta} [ \left(\epsilon+P_L\right)
   - \frac{1}{\tau v_{\eta}} \left(1+\tau^2 v^2_{\eta}\right)q ].
\label{evo4}
\end{align}
Here we have used the shorthand notation $\tilde{A}^{mn}=\tau A^{mn}$, 
with $P_L=P-\pi$ as the effective longitudinal pressure and 
$\tilde{v}_\eta=\tilde{T}^{\tau\eta}/\tilde{T}^{\tau\tau}=T^{\tau\eta}/T^{\tau\tau}$.
The equation for number density (\ref{thid1}) simply turns out to be $Dn+n\theta=0$.

Using Eqs. (\ref{evo3}) and (\ref{evo4}), the expression for energy density and longitudinal velocity 
can be expressed as 
\begin{align}
\epsilon & =  T^{\tau\tau}-\tau^2 v_{\eta} T^{\tau\eta}+\tau v_{\eta}q,\\
v_{\eta} & = \frac{T^{\tau\eta} + q/\tau}{T^{\tau\tau} 
+ P(\epsilon = T^{\tau\tau}-\tau^2 v_{\eta} T^{\tau\eta}+\tau v_{\eta}q) - \pi}.
\end{align}
This allows to extract $v_\eta$ by one-dimensional zero-search.
The five evolution equations (\ref{evo1}), (\ref{evo2}), (\ref{evo3}), (\ref{evo4}), and
the baryon number evolution,
in terms of six unknowns ($\pi, q, v_{\eta},\epsilon$, $P$ and $n$),
are closed by including an equation of state  $P=P(n,\epsilon)$.
The above set of evolution equations are solved using SHASTA-FCT algorithm.

For the (1+1)D flow, the entropy density and the components of the entropy flux reduce to 
\begin{align}
& s=s_{eq} -  \frac{1}{(\epsilon+P)T} \left(\frac{9}{4}\pi^2 + 3 q^2\right), \label{evoent1}\\
& \Phi^{\eta}= -\gamma\Phi/\tau , ~~~ \Phi^{\tau}=-\gamma\tau v_{\eta}\Phi, \label{evoent2}
\end{align}
with $\Phi=(q/T)[1+2\pi/(\epsilon+P)]$.

\section{Equation of state and freeze-out}

To study the hydrodynamic evolution of matter at finite density, we have employed the (2+1)-flavor 
QCD EoS where the bulk thermodynamic quantities have been obtained by using Taylor series expansion 
up to sixth order in the baryon chemical potential \cite{Bazavov}. The expansion was constructed about
the lattice QCD EoS at $\mu_B=0$ that includes a crossover transition at $T_c \simeq 154$ MeV \cite{Bazavov:2014pvz}
and a hadron resonance gas (HRG) EoS at lower $T$. The parametric EoS was shown to be reliable at $0 < \mu_B/T < 2$, 
and hence, can be suitably used about midrapidity for the beam energy scan $\sqrt{s_{NN}} \sim 5 - 200$ GeV.
At $T < 130$ MeV and/or high baryon density encountered at the forward/backward rapidities, where the 
Taylor expansion is not well-defined, we have used the HRG EoS smoothly matched to the parametric EoS.

The hadron spectra at freeze-out can be obtained by using the standard Cooper-Frye prescription \cite{CooperFrye},
\be
E\frac{dN_i}{d^3p}=\frac{g_i}{(2\pi)^3} \int_{\Sigma} d\Sigma_\mu \: p^\mu f^i(x,p).
\label{CF}
\ee
We have considered freeze-out at a constant decoupling temperature of $T_{\rm dec}$ 
(that corresponds to freeze-out times $\tau_f(\eta)$) from the hypersurface $\Sigma (x)$. 
The phase-space distribution function at freeze-out can be expressed as 
$f^i(x,p) = f^i_0(x,p) + \delta f^i(x,p)$. The local equilibrium distribution is
$f^i_0 = [\exp(u^\mu p_\mu - b_i \mu_B)/T \pm 1]^{-1}$
where $b_i$ is the baryon number for $i$th species and the flow velocity $u^\mu \equiv u^\mu(\tau_f,\eta)$,
temperature $T \equiv T(\tau_f,\eta)$ and baryon chemical potential $\mu_B \equiv \mu_B(\tau_f,\eta)$ 
are evaluated at the freeze-out hypersurface coordinates. The nonequilibrium corrections $\delta f$ from shear
viscosity and thermal conduction, obtained in Eq. (\ref{Gradphi1}) from Grad's approach, can be written as
\begin{align}
\delta f^i & = \delta f^i_{\rm visc} + \delta f^i_{\rm ther} = f^i_0(1 \pm f^i_0) \nn
& \times \left[ \frac{p^\mu p^\nu \pi_{\mu\nu}}{2 T^2(\epsilon +P)}
+ \frac{p^\mu q_\mu}{T(\epsilon +P)} \left( \frac{(p^\nu u_\nu - 4T)n_B}{(\epsilon+P) - 5n_BT} \right) \right].
\end{align}
We have considered all the resonances that are used in the HRG model, and the rapidity distribution 
presented here include two- and three-body resonance decays \cite{Bhalerao:2015iya}.

\section{Numerical results}

We will present (1+1)D hydrodynamic simulation results over the entire rapidity range
corresponding to Au+Au collisions at $\sqrt{s_{NN}}= 200$ GeV at RHIC, 
Pb+Pb collisions at $\sqrt{s_{NN}}= 17.3$ GeV at SPS/CERN, and Au+Au collisions at $\sqrt{s_{NN}}= 5$ GeV 
at AGS/BNL that is close to the lowest energy in RHIC BES. We have taken the initial time as $\tau_0=0.4$ fm 
at which the initial energy density profile is constructed as \cite{Chattopadhyay:2018dth}
\begin{align}\label{engd:eq}
\epsilon(\tau_0,\eta) = \epsilon_0 \exp\left[ - \frac{\left(|\eta| - \Delta\eta_\epsilon\right)^2}
{2\sigma_{\eta_\epsilon}^2} \theta\left(|\eta| - \Delta\eta_\epsilon\right) \right],
\end{align}
which consists of a flat distribution about midrapidity of width $2\Delta\eta$  and two-smoothly
connected Gaussian tails of half-width $\sigma_{\eta_\epsilon}$. The initial net-baryon density 
profile is taken as $n_B(\tau_0,\eta) \equiv n_{B_0} \: f_B (\eta)$, where the envelope function
is chosen to be 
\begin{align}\label{bprof:eq}
f_B (\eta) & = \theta\left(|\eta| - \eta_B^\pm \right)
\exp \left(- \frac{\left(|\eta| - \eta_B^\pm \right)^2}{2\sigma_{\eta>}^2} \right) \nonumber \\
& + \theta\left(\eta_B^\pm  - |\eta| \right) \Bigg[ 
\exp \left(- \frac{\left(|\eta| - \eta_B^\pm \right)^2}{2\sigma_{\eta<}^2} \right) 
 \theta\left(|\eta| - \Delta\eta_{B_0}\right) \nonumber \\
& ~~~~~~~~~~~~~ \ + {\cal F}(\Delta\eta_{B_0}) \left(\Delta\eta_{B_0} - |\eta|\right) \Bigg].
\end{align}
This represents a flat distribution about midrapidity of width $2\Delta\eta_{B_0}$  which
is connected smoothly at $\pm \Delta\eta_{B_0}$ to the tails of two Gaussian with
width parameter $\sigma_{\eta<}$ and peak position $\eta_B^\pm$ that are determined from
the measured shape of the final net-proton rapidity distribution and the rapidity-loss
in the net-proton distribution, respectively \cite{Denicol:2018wdp}.
The initial net-baryon envelope profile at RHIC is shown in Fig. \ref{dndy:fig} (dotted line).

The energy and baryon profiles of Eqs. (\ref{engd:eq}) and (\ref{bprof:eq}) are relevant at the 
RHIC and SPS energies that produce a boost-invariant matter with small net-baryon density at 
midrapidity. In contrast, at $\sqrt{s_{NN}} \approx 5$ GeV,
complete stopping of the colliding nuclei leads to a baryon-rich and high-energy density matter at 
midrapidity, and hence require (initial) Gaussian profiles peaked at $\eta=0$.
The parameters in Eqs. ({\ref{engd:eq}) and (\ref{bprof:eq}) are adjusted to reproduce the
net-proton and charged pion rapidity distributions $dN/dy$ at RHIC, SPS and AGS. 
The initial values of the longitudinal velocity profile is taken as
boost-invariant, viscous stress tensor as isotropic, and vanishing heat flow, i.e.
$v_\eta(\tau_0,\eta) = 0, ~\pi^{mn}(\tau_0,\eta) = 0, ~q^\mu(\tau_0,\eta) = 0$.

%%%%%%%%%%%%%%%%%%%%%%%%%%%%%%%%%%%%%%%%%
\begin{figure}[t]
\includegraphics[scale=0.42]{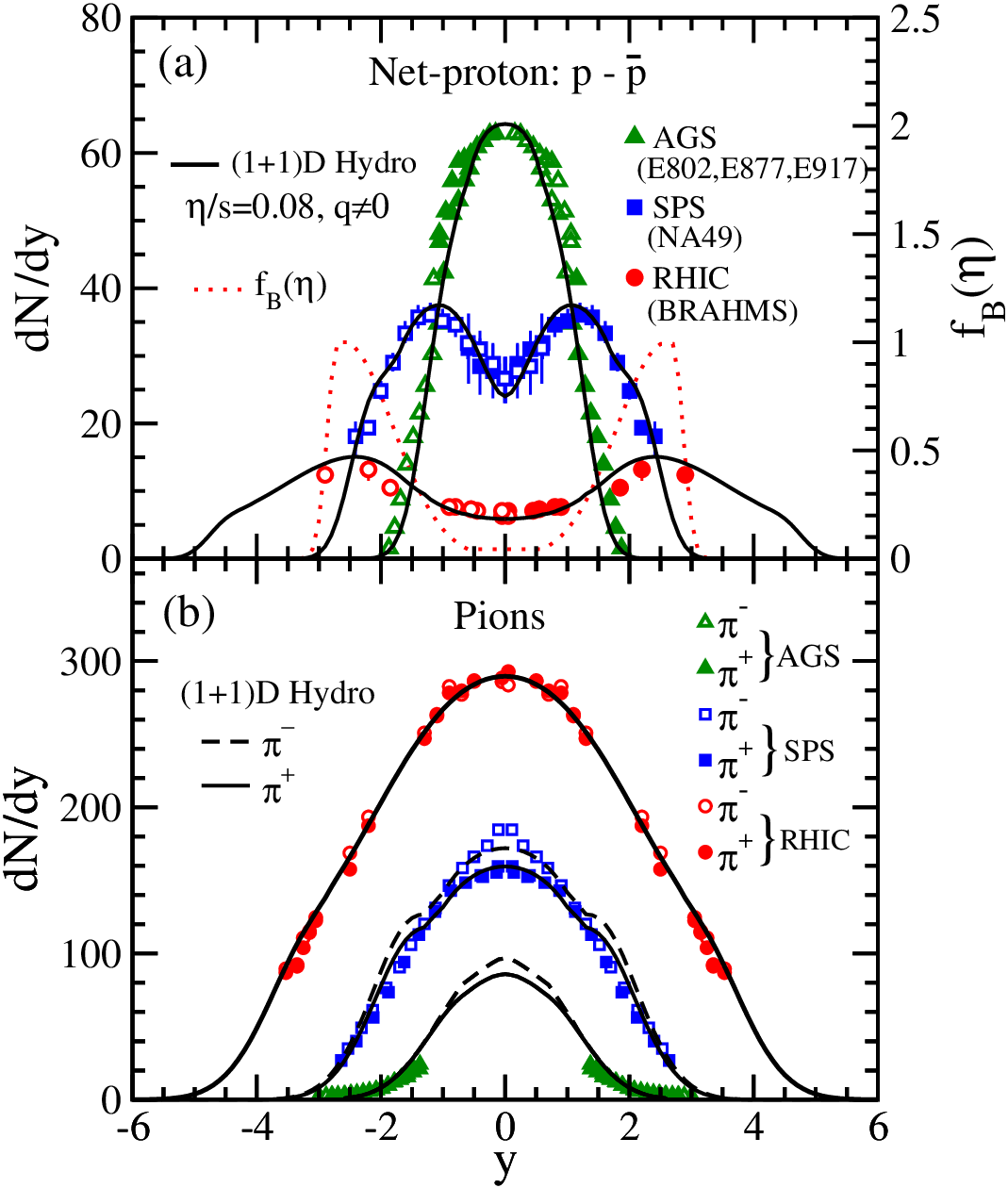}
\caption{Rapidity dependence of net-protons (top panel) and charged pions (bottom panel) in heavy ion
collisions at the RHIC energy $\sqrt{s_{NN}}=200$ GeV, at the SPS energy $\sqrt{s_{NN}}=17.3$ GeV, and at
the AGS energy $\sqrt{s_{NN}}=5$ GeV. The lines correspond to non-boost-invariant hydrodynamic 
calculations with shear stress and heat flow using $\eta_v/s=1/4\pi$ and Eq. (\ref{lambda:eq}).
The symbols represent $dN/dy$ data at AGS \cite{E877-ppi}, SPS \cite{NA49-p,NA49-pi},
and RHIC \cite{BRAHMS-p,BRAHMS-pi}. The dotted line (top panel) illustrates the initial net-baryon 
density envelope profile at RHIC.}
\label{dndy:fig}
\end{figure}
%%%%%%%%%%%%%%%%%%%%%%%%%%%%%%%%%%%%%%%%%

For QCD matter at high $T$, the thermal conductivity $\lambda$ can be estimated from kinetic
theory in the relaxation time approximation as \cite{Hosoya,Kapusta-Rincon}
\begin{align}\label{lambda:eq}
\lambda & = \frac{1}{3T^2}\int \frac{d{\bf p}}{(2\pi)^3} \: \Bigg[ |{\bf p}|^2 g_b\tau_b(p) b_0(1+b_0)\nonumber\\
& + \sum_{i=f,\bar f}  g_i\tau_i(p) f_{i0}(1-f_{i0})\left(\frac{{\bf p}}{\epsilon_i} \right)^2 
\left(\epsilon_i - b_i\frac{\epsilon+P}{n_B}\right)^2 \Bigg] , 
\end{align}
which includes bosons and massless quarks/antiquarks ($b_i = \pm 1$) 
with single-particle energy $\epsilon_i = \sqrt{m_i^2 + {\bf p}^2}$.
We have used thermally averaged values for relaxation times $\tau_i(p)$ from \cite{Hosoya}.
In the hadronic phase at low $T$, the thermal conductivity is obtained from Eq. (\ref{lambda:eq}) 
by summing over all the (anti-)baryons $i$ with baryon number $b_i$.
Throughout our analysis, we have taken a constant value of shear viscosity to entropy density ratio 
of $\eta_v/s = 0.08$.

In Fig. (\ref{dndy:fig}) we show the net-proton (top panel) and charged pion (bottom panel) rapidity distribution
in (1+1)D hydrodynamics that include both viscous and heat dissipation and compare with the $dN/dy$ data at
RHIC, SPS and AGS. Contribution from resonance decays are added to the thermal distribution. For the good
description of the $dN/dy$ data, we require an initial $\epsilon_0(\tau_0) = 23.5, 16.2, 9.0$ GeV/fm$^3$, 
and a decoupling temperature of $T_{\rm dec} = 150, 142, 135$ MeV for RHIC to AGS energies.
Further, a small $n_B(\tau_0)$ is required in the boost-invariant central rapidity region at RHIC (dotted curve), 
whose magnitude gradually increases and extent decreases at smaller colliding energies.
The observed $dN/dy$ at RHIC is the culmination of near cancellation of proton and antiproton values 
near $y \approx 0$ and a doubled peaked structure for the proton $dN/dy$ at the large rapidities having 
high baryon content.

In the present analysis, with and without viscous and heat dissipations, the parameters in 
$\epsilon(\tau_0,\eta)$ are retuned in each case to fit the hadron $dN/dy$.
As compared to dissipative hydrodynamics, the larger flow in the ideal fluid 
requires a higher and slightly narrower initial energy
density distribution to be compatible with the final $dN/dy$.

%%%%%%%%%%%%%%%%%%%%%%%%%%%%%%%%%%%%%%%%%
\begin{figure}[t]
\includegraphics[scale=0.35]{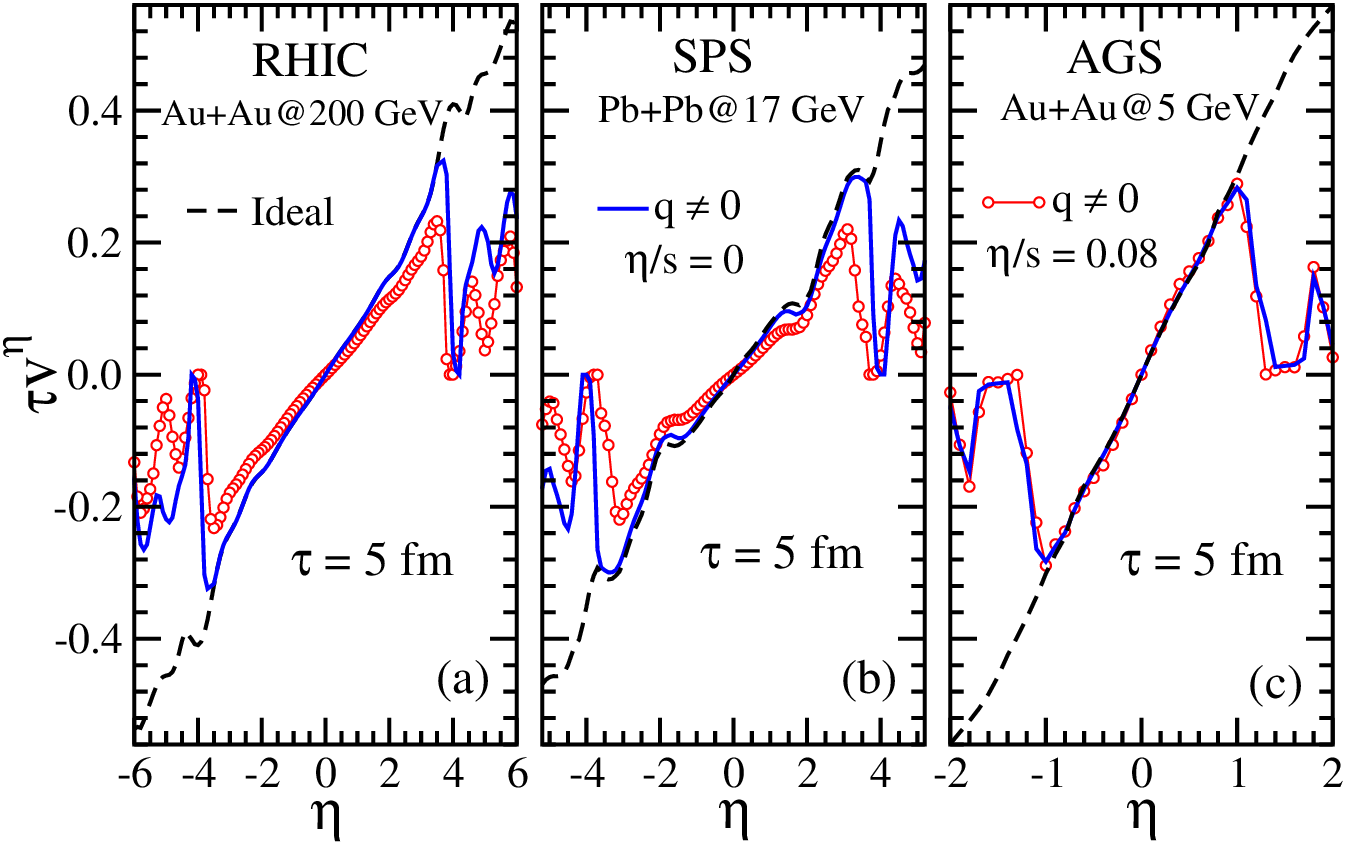}
\caption{Space-time rapidity dependence of longitudinal velocity (scaled by time) in non-boost-invariant
hydrodynamic simulation for ideal flow (dashed lines), with inclusion of heat flow (solid lines) and further
inclusion of shear stress (symbols) in heavy ion collisions at RHIC, SPS, AGS energies.}
\label{vel:fig}
\end{figure}
%%%%%%%%%%%%%%%%%%%%%%%%%%%%%%%%%%%%%%%%%

Figure (\ref{vel:fig}) shows the space-time rapidity dependence of the longitudinal flow velocity $v_{\eta}$ 
(multiplied by the corresponding proper time) in ideal, pure thermal, 
and thermal plus viscous hydrodynamics at $\tau=5$ fm. Large pressure gradients $P_L$, 
in the (1+1)D ideal-fluid-expansion, breaks the initial boost-invariance $v_{\eta}(\eta,\tau_0)=0$ 
and accelerates the fluid flow towards high rapidity. Shear and heat dissipation
will restrict $P_L$, and finally overcome it at large $\eta$ where
small ($\epsilon,P,T$) and large $n_B$ increase the times,
$\tau_\pi  \sim \eta_v/(\epsilon+P)$ and $\tau_q  \sim \lambda T n_B/(\epsilon+P)$, for 
the system to relax to equilibrium.  We find that at RHIC and SPS energies,
viscous drag is more effective than thermal correction as $P_L= P-\pi$. 
Whereas, for collisions at the low $\sqrt{s_{NN}} \sim 5$ GeV, the
viscous effects are rather small, and sizable effects from baryon and temperature gradients 
on the heat flow modify $v_\eta$ close to midrapidity.
The oscillations in $v_\eta$ seen at larger rapidities stem for large gradients in
$\mu_B/T$ near the vacuum \cite{Li-Shen}.

%%%%%%%%%%%%%%%%%%%%%%%%%%%%%%%%%%%%%%%%%
\begin{figure}[t]
\includegraphics[scale=0.42]{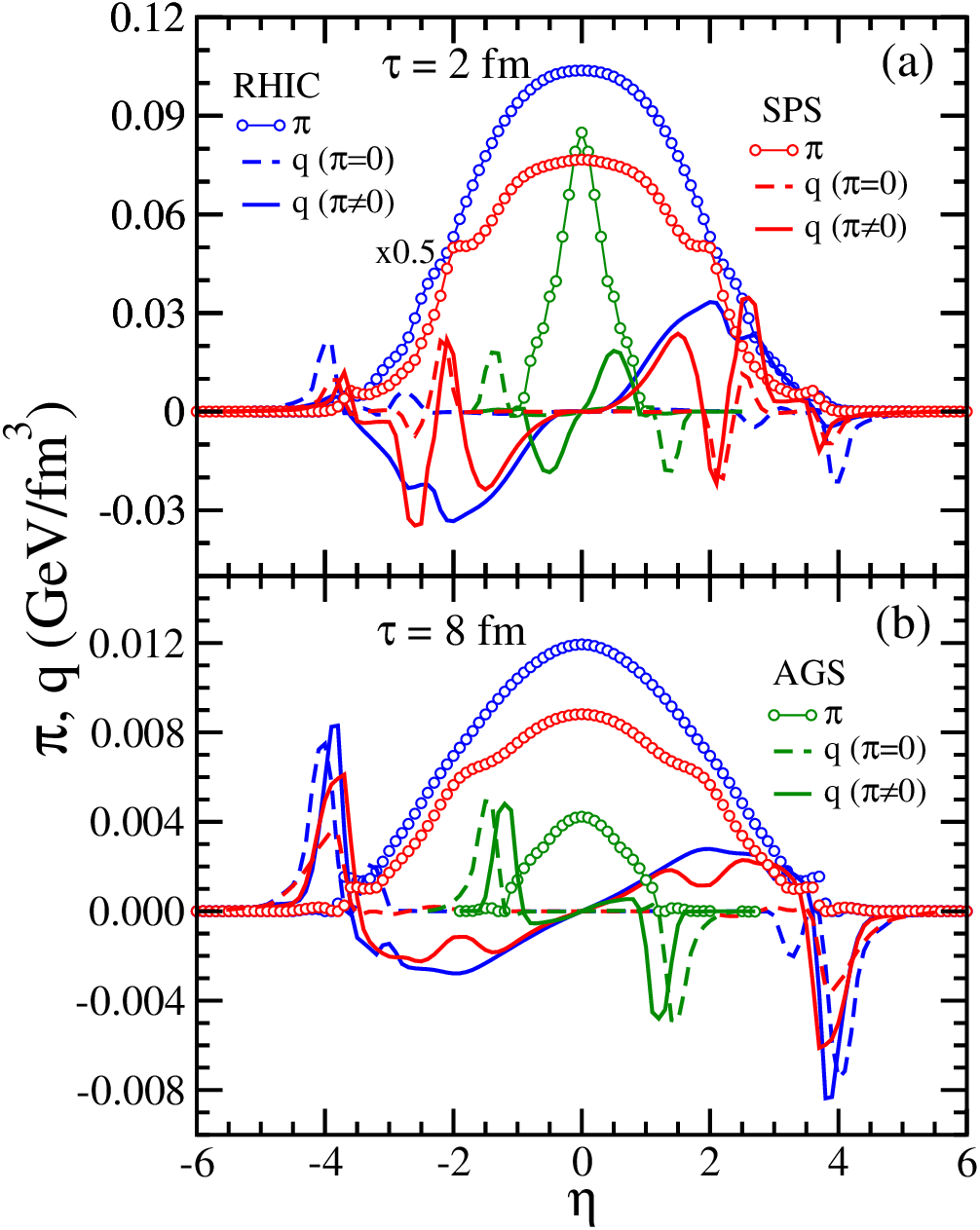}
\caption{Space-time rapidity dependence of shear pressure tensor $\pi$ and heat flow $q$ at proper time
$\tau=2$ (top panel) and 8 fm (bottom-panel) at the RHIC, SPS and AGS energies 
in the (1+1)D hydrodynamic calculations that include heat flow in absence of shear (dashed lines),
with further inclusion shear (solid lines), and for shear stress (symbols).} 
\label{heatsh:fig}
\end{figure}
%%%%%%%%%%%%%%%%%%%%%%%%%%%%%%%%%%%%%%%%%

Figure (\ref{heatsh:fig}) shows the $\eta$ dependence of shear pressure $\pi$ and heat flow $q$ 
at RHIC, SPS and AGS energies at $\tau=2$ and 8 fm. In absence of shear, 
the magnitude of the heat flow (dashed lines) 
at the central $\eta$ does not build up with time from its initial value of 
$q(\tau_0,\eta) = 0$. This arises due to small $|v_\eta|$ 
and near cancellation of $(T,P)$ gradients at small $\eta$ in the dominant first term
in Eq. (\ref{evo2}). The peaks and oscillations at large rapidities near the vacuum can be traced to rapid heat 
flow in the direction of large gradients in $\mu_B/T$ \cite{Li-Shen}.

On inclusion of shear, the magnitude of heat current (solid lines) increases considerably 
from heat-shear coupling, and shows a trend similar to $v_\eta$ (see Eq. (\ref{evo2}) and Fig. \ref{vel:fig}). 
At the early time $\tau=2$ fm, the heat flow at RHIC and SPS has a vanishingly 
small value at $|\eta| < 1$  due to the initial boost-invariance, 
and a maximum value near the nuclear remnants. In contrast, at lower c.m. energies, 
the large gradients in $n_B,P,T$ at small $\eta$ cause a rapid increase in $q$; the enhancement 
here is however smaller due to small shear. In any case, the magnitude of $q$ is expected to be much smaller 
\cite{Bouras} than the shear stress (symbols). 

As the system evolves to a later time $\tau=8$ fm, the typical features seen at the RHIC and SPS are 
breaking of boost-invariance and inward-outward baryon diffusion from high rapidities. This causes a reduction
in central plateau and broadening of the peak for heat flow. In contrast at AGS, the
baryons rapidly diffuses out from the midrapidity resulting in heat to flow out 
from central to larger rapidities. Close inspection of Fig. (\ref{heatsh:fig})(a) and 
(\ref{heatsh:fig})(b) reveals that, with the expansion of the system, the shear pressure drops faster as compared 
to the heat flow.

%%%%%%%%%%%%%%%%%%%%%%%%%%%%%%%%%%%%%%%%%
\begin{figure}[t]
\includegraphics[scale=0.40]{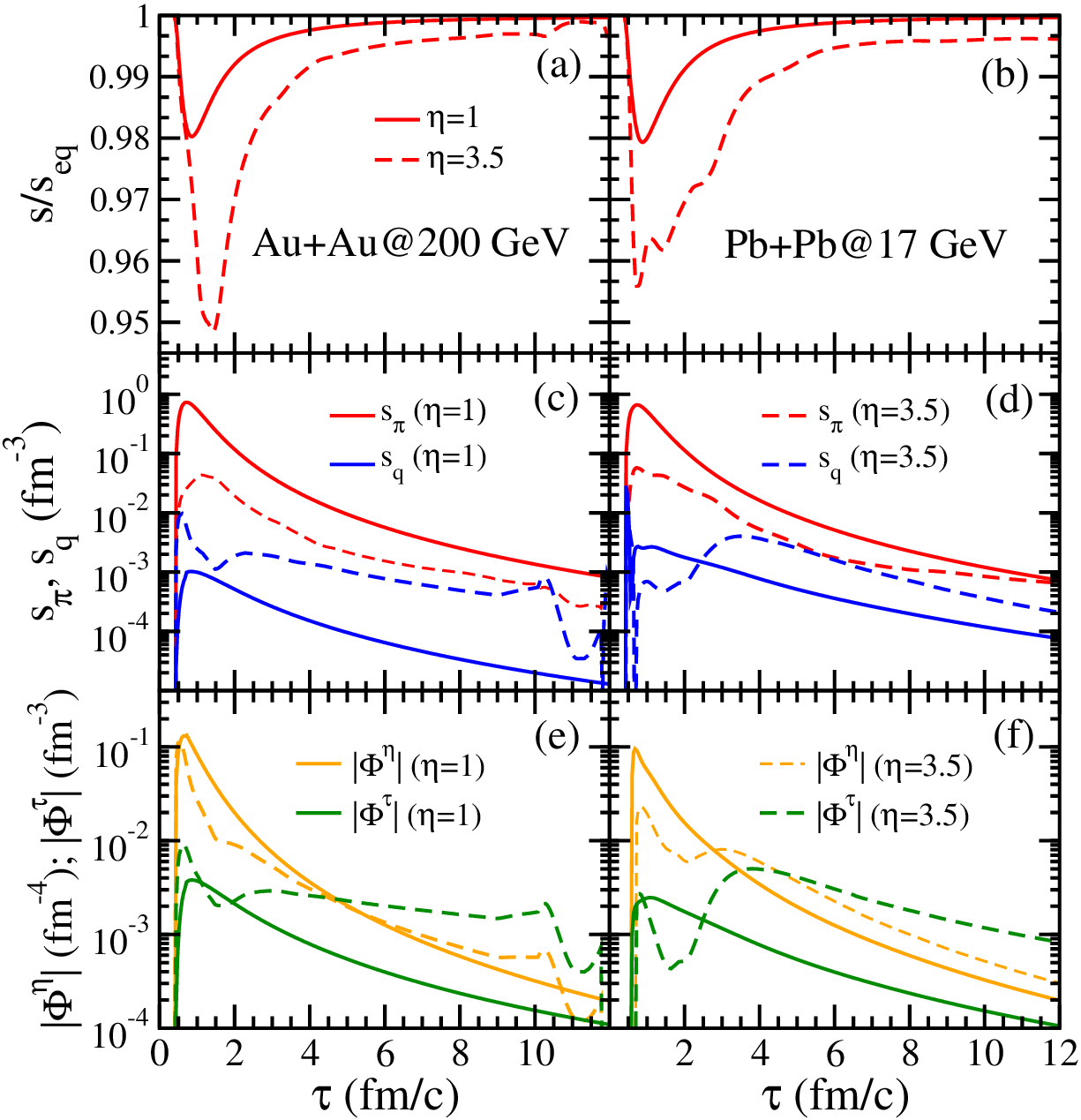}
\caption{Time evolution of normalized entropy density $s/s_{\rm eq}$ (top panels), entropy density from viscous $s_\pi$
and heat $s_q$ (middle panels), and longitudinal and temporal components of entropy flux
$s^\eta_q$ and $s^\tau_q$ (bottom panels) at the space-time rapidity $\eta=1$ (solid lines) 
and $\eta=3.5$ (dashed lines) in heavy-ion collisions at RHIC (left panels) and SPS (right panels).}
\label{entropy:fig}
\end{figure} 
%%%%%%%%%%%%%%%%%%%%%%%%%%%%%%%%%%%%%%%%%

Figure \ref{entropy:fig}(a)-(f) displays the proper time evolution of various nonequilibrium components of 
entropy density from Eq. (\ref{evoent1}) at the RHIC (left panels) and SPS (right panels) for 
$\eta =1$ (solid lines) and $\eta=3.5$ (dashed lines). 
The normalized entropy density $s/s_{eq}$ in Figs. \ref{entropy:fig}(a)-(b) 
represent the deviation from the equilibrium value $s_{\rm eq} = (\epsilon + P -\mu n_B)/T$ 
due to thermal and viscous dissipation. Such a time dependence essentially reflect the 
evolution of normalized shear tensor $\pi/(\epsilon+P)$ and heat flow $q/(\epsilon+P)$ that
rapidly increases from the initial value of $\pi(\tau_0)= q(\tau_0)=0$ and then gradually decreases.
At a larger rapidity $\eta=3.5$ and also at the lower SPS energy, the nonequilibrium deviation is found 
to be enhanced due to the growth of the relaxation times, $\tau_\pi$ and $\tau_q$, that causes a slower 
response to the expansion, driving the system away from equilibrium.

Figures \ref{entropy:fig}(c)-(d) depict the viscous 
$s_{\pi}=9\pi^2/4T(\epsilon+P)$ and thermal $s_q= 3q^2/T(\epsilon+P)$ contributions to the entropy density
which again reflect the time evolution seen for shear and heat.
While $s_{\rm \pi}$ is larger at central than at higher rapidities, 
$s_q$ shows an opposite $\eta$-dependence due to relatively
larger $|q|$ at the nuclear spectator regions. We find that $s_q$ and $s_\pi$ are
of comparable magnitude at high rapidities.

Finally, we show in Figs. \ref{entropy:fig}(e)-(f) the magnitude of 
longitudinal $s^\eta_q$ and temporal $s^\tau_q$ components of entropy flux of Eq. (\ref{evoent2}). 
These are the signature of thermodynamic quantities originating purely from the heat flow.
Apart from the general decreasing trend with time, both these components 
dominate at large rapidity most of the time during evolution,
demonstrating the key role played by thermal flux $q$ in deciding the behavior of entropy flux contributions.

\section{Summary and conclusion}

We have presented a relativistic hydrodynamic formulation, with thermal conductivity and shear viscosity, to treat
high temperature and finite net-baryon density matter encountered in relativistic heavy ion collisions.
The dissipative equations obtained within Grad's 14-moment approach have new terms involving couplings
between heat flow and shear pressure where all the second-order transport coefficients are explicitly determined.
Numerical significance of these equations were explored for non-boost-invariant expansion 
of matter created in heavy ion collisions over a wide energy range.   
We have employed a (2+1)-flavor QCD EoS using up to sixth-order Taylor expansion 
in baryon chemical potential and smoothly matched to the HRG EoS at low temperature and density. 
We found that thermal conduction alone has a small effect in the central region at RHIC and SPS 
but has noticeable effects in the baryon-rich regions and in lower energy heavy-ion collisions. 
Coupling to shear pressure could result in a sizable contribution from heat flow  on the observables.
The present study may be promising in the search for the elusive critical point at finite density 
and temperature in the QCD phase diagram.

\end{document}